\documentclass{aastex6}

\begin{document}


\title{Temporary Capture of Asteroids by an Eccentric Planet}


\author{A. Higuchi}
\affil{Department of Earth and Planetary Sciences, Faculty of Science, 
  Tokyo Institute of Technology, Meguro, Tokyo 152-8551, Japan}
\and
\author{S. Ida}
\affil{Earth-Life Science Institute,
  Tokyo Institute of Technology, Meguro, Tokyo 152-8550, Japan}

\begin{abstract}
  We have investigated the probability of temporary capture of asteroids in eccentric orbits
  by a planet in a circular or an eccentric orbit through analytical and numerical calculations.
  We found that in the limit of the circular orbit,
  the capture probability is $\sim 0.1\%$ of encounters to the planet's Hill sphere,
  independent of planetary mass and semimajor axis.
  In general, the temporary capture becomes more difficult as the planet's eccentricity ($e_{\rm p}$) increases.
  We found that the capture probability is almost independent of $e_{\rm p}$ until 
  a critical value ($e_{\rm p}^{\rm c}$) that is given by $\simeq$ 5 times the Hill radius scaled by the planet's semimajor axis.
  For $e_{\rm p} > e_{\rm p}^{\rm c}$, the probability decreases
  approximately in proportion to $e_{\rm p}^{-1}$.
  The current orbital eccentricity of Mars is several times larger than $e_{\rm p}^{\rm c}$.
  However, since the range of secular change in Martian eccentricity overlaps $e_{\rm p}^{\rm c}$,
  the capture of minor bodies by the past Mars is not ruled out.
\end{abstract}



\section{Introduction}

  Irregular satellites around giant planets, which are small
  and with elliptical and inclined orbits,
  are usually thought to be captured passing asteroids
  \citep[e.g.,][]{jh07, n08}.
The objects captured temporarily in the Hill sphere of a planet can be 
  permanently captured
  by some energy loss (e.g., tidal dissipation, 
  drag force from a circumplanetary disk when it existed,
  or collisions with other solid bodies in the disk).
 \citet{hi16} derived the conditions for the temporary capture
  by a planet in a {\it circular} orbit 
  as functions of the mass and semimajor axis of the host planet,
  and clarified the range of semimajor axes
  of field particles for prograde and retrograde capture.

\citet{hi16} commented that the small eccentricity of Jupiter does not
  affect the capture probability.
  However, the effect of a high eccentricity like that of Mars has not been
  investigated.
  Mars has two satellites: Phobos and Deimos.
  Two major theories of the origin of these satellites are
  (1) in situ formation through accretion of an impact-generated debris 
   by a large impact inferred from the Borealis basin 
  \citep[e.g.,][]{c15,r16}
  and (2) capture of asteroids \citep[e.g.,][]{b78}.
  While the large impact model may explain the circular, non-inclined orbits of Phobos and Deimos,
  which is not easily explained by the capture origin, 
  the surface characteristics of the satellites are similar to those of primitive asteroids.
  Spectral observations of Phobos and Deimos suggest that
  the material of the satellites is best modeled as a primitive material,
  which may not be easily explained by the large impact origin \citep{f14}.
  Future sample return missions, such as MMX (Mars Moon eXploration), will 
  provide important clues
  about the Martian satellite origin.
  It is important to explore the possibility of the capture origin model in detail, as well as to investigate
the large impact model.

  In this study, we generalize our previous study to investigate the effects of orbital eccentricity
  of a planet on the temporary capture probability through analytical and numerical calculations.
  We derive the probability of temporary capture from encounters with the planet's Hill sphere
  as a function of planetary eccentricity $e_{\rm p}$ and mass $m_{\rm p}$.
  If the encounter frequency is given by other simulations, we can evaluate the probability of temporary capture throughout 
  the history of the solar system. 
  Most of our analysis and orbital calculations assume planar orbits,
  but some calculations are done with small finite inclinations.
    Temporary capture is a necessary condition for permanent capture.
    The relation between temporary and permanent capture will
    be investigated in a subsequent paper.

  We summarize the assumptions, basic formulation, and derivation
  of the analytical formulae in Section 2.
  We define and derive the efficiency of temporary capture in Section 3.
  The methods and results of numerical calculations are presented
  and compared with the analytical prediction in Section 4.
  In Section 5, we summarize the results and comment on the origin of Martian satellites.

\section{ANALYTICAL DERIVATION OF TEMPORARY CAPTURE EFFICIENCY BY AN ECCENTRIC PLANET}

We first derive analytical formulae for temporary capture
by an eccentric planet.
These formulae give orbital elements
of the asteroids that can be captured, as functions of
 the mass, eccentricity, and true anomaly of the host planet.
 As we will show in section 4, the analytical formulae reproduce the results obtained through
 numerical orbital integrations.
From the analytical derivation, the intrinsic dynamics of temporary capture by an eccentric planet will be revealed.

\subsection{Assumptions}
Following \citet{hi16}, we split a coplanar three-body problem (Sun-planet-particle) into 
 two independent two-body problems (Sun-particle and planet-particle).
The particles are candidates that are captured by the planet to become satellites.
 Hereafter, we refer to particles as "asteroids," although the particles do not necessarily originate from the asteroid belt. 
We identify the relative velocity between the asteroid and the planet
in heliocentric orbits with the satellite velocity orbiting around the planet 
 (condition [1]) at the capture point.
The capture points are assumed to be the $L_1$ and $L_2$ points (condition [2]).
The distance of the points from the planet is the Hill radius, 
\begin{eqnarray}
  r_{\rm H}  = r_{\rm p} \left(\frac{m_{\rm p}}{3M_\odot}\right)^{\frac{1}{3}} =  r_{\rm p}\hat{r}_{\rm H},
\label{eq:ep_rs}
\end{eqnarray}
where  $m_{\rm p}$ is the planet mass and $M_\odot$ is the solar mass.
The instantaneous heliocentric distance is given by 
\begin{eqnarray}
  r_{\rm p}&=&a_{\rm p}\frac{1-e_{\rm p}^2}{1+e_{\rm p}\cos f_{\rm p}},
  \label{eq:ep_rp}
 \end{eqnarray}
where $a_{\rm p}$, $m_{\rm p}$, $e_{\rm p}$, and $f_{\rm p}$
are semimajor axis, mass, eccentricity, and true anomaly
of the planet, respectively.
We also assume that the geometric condition that the two elliptic orbits are touching at
 a capture point; the velocity vectors of the planet
and the asteroid are parallel or antiparallel (condition [3]).

\subsection{Conditions for Temporary Capture}

We consider an asteroid and a planet in the 
Cartesian coordinates
($x$, $y$) centered on the Sun.
The $x$-axis is toward the perihelion of the planet's orbit
and the $x$-$y$ plane lies in the planet's orbital plane.
Let $r$, $a$, $e$, and $f$ be  heliocentric distance,
semimajor axis, eccentricity of the asteroid,
and true anomaly, respectively.

Condition [2] reads as
\begin{eqnarray}
  r = a\frac{1-e^2}{1+e\cos(f+\theta)}   =  r_{\rm p}A_\mp,
  \label{eq:ep_rr0}
  \label{eq:ep_rr}
\end{eqnarray}
where $\theta=f_{\rm p} - f$ and 
\begin{eqnarray}
  \left\{
  \begin{array}{l}
    A_- = 1-\hat{r}_{\rm H}\;\;{\rm at}\;\;L_1 \\
    A_+ = 1+\hat{r}_{\rm H}\;\;{\rm at}\;\;L_2.
  \end{array}
  \right.
  \label{eq:ep_aa}
\end{eqnarray}
The heliocentric velocity of the asteroid at capture is
\begin{eqnarray}
  v = \sqrt{GM_\odot\left(\frac{2}{r}-\frac{1}{a}\right)} = \sqrt{GM_\odot\left(\frac{2}{A_\mp r_{\rm p}}-\frac{1}{a}\right)}
    = v_{\rm p}\chi, 
 \label{eq:ep_v2}
  \label{eq:tc_v}
\end{eqnarray}
where $v_{\rm p}=\sqrt{GM_\odot/a_{\rm p}}$ and
\begin{eqnarray}
    \chi =  \sqrt{\frac{2}{\Phi_{\rm p}A_\mp}-\frac{a_{\rm p}}{a}},
    \label{eq:ep_chi}
    \\
    \Phi_{\rm p}=\frac{1-e_{\rm p}^2}{1+e_{\rm p}\cos f_{\rm p}}.
    \label{eq:ep_phip}
\end{eqnarray}

Condition [1] reads as
\begin{eqnarray}
  {\bf v}-{\bf v}_{\rm p} = {\bf v}_{\rm s},
  \label{eq:ep_vr}
\end{eqnarray}
where  ${\bf v}$ and ${\bf v}_{\rm p}$ are heliocentric velocities of the asteroid
and the planet and ${\bf v}_{\rm s}$ is the planetocentric velocity of the asteroid as a satellite at the capture. 
The velocity of the satellite at the planetocentric distance $r_{\rm s}=r_{\rm H}$ is 
\begin{eqnarray}
  v_{\rm s} = \sqrt{Gm_{\rm p}\left(\frac{2}{r_{\rm H}}-\frac{1}{a_{\rm s}}\right)}
  = v_{\rm H} \sqrt{2 - \Phi_{\rm s}},
  \label{eq:ep_vs}
\end{eqnarray}
where $a_{\rm s}=r_{\rm H}/\Phi_{\rm s}$ is the planetocentric semimajor axis of the satellite, 
\begin{eqnarray}
  \Phi_{\rm s}&=&\frac{1-e_{\rm s}^2}{1+e_{\rm s}\cos f_{\rm s}}, \label{eq:ep_phis} \\
  v_{\rm H}&=&\sqrt{\frac{Gm_{\rm p}}{r_{\rm H}}} =\sqrt{\frac{3}{\Phi_{\rm p}}}\hat{r}_{\rm H},
  \label{eq:ep_vss}
\end{eqnarray}
and $e_{\rm s}$ and $f_{\rm s}$ are the planetocentric eccentricity and true anomaly.
Since $ v_{\rm H}$ is a circular velocity around the planet at the
 planetocentric distance $r_{\rm s}=r_{\rm H}$, 
\begin{eqnarray}
  \nu = v_{\rm s}/v_{\rm H} =  \sqrt{2 - \Phi_{\rm s}}
  \label{eq:ep_vs2}
\end{eqnarray}
is related to the planetocentric orbital eccentricity (which is 
equivalent to $\kappa^2$ appearing in \citet{hi16}); $\nu=1$ corresponds to a circular orbit with the semimajor axis
$a_{\rm s}=r_{\rm H}$ and the orbit is hyperbolic for $\nu>\sqrt{2}$.

Condition [3] is expressed by $\alpha=\alpha_{\rm p}$, where
$\alpha$ and $\alpha_{\rm p}$ are the angles between the position and
 velocity vectors of the asteroid and those of the planet, which are given by
 \begin{eqnarray}
  \sin\alpha&=&\frac{1+e\cos(f+\theta)}{\sqrt{1+e^2+2e\cos(f+\theta)}}
  \\
  \sin\alpha_{\rm p}&=&\frac{1+e_{\rm p}\cos f_{\rm p}}{\sqrt{1+e_{\rm p}^2+2e_{\rm p}\cos f_{\rm p}}}
\end{eqnarray}
These angles are given geometrically, applying the law of cosines to
 a triangle composed of ${\bf r}$, the $x$ axis, and the tangent line of the orbit at $r$. 
Another way to derive $\alpha$ using the angular momentum is found in \citet{roy05}.

\subsection{Equation of Temporary Capture}
We combine the equations describing the three conditions above
and solve for the orbital elements of temporarily captured asteroids.

\subsubsection{Derivation of Heliocentric Orbital Elements for Temporary Capture}
\paragraph{Semimajor axis.}
Using ${\bf v}\parallel{\bf v}_{\rm p}$ (condition [3]), $v_{\rm s} = \nu v_{\rm H}$
(Eq.~\ref{eq:ep_vs}),  and  $v = v_{\rm p} \chi$ (Eq.~\ref{eq:ep_vs2}),
condition [1] (Eq.~\ref{eq:ep_vr}) becomes
\begin{eqnarray}
  |\chi-1| =    \nu v_{\rm H}.
  \label{eq:ep_vrs}
\end{eqnarray}
Substituting Equations (\ref{eq:ep_rr}),(\ref{eq:ep_aa}), and
(\ref{eq:ep_chi}), 
into Equation (\ref{eq:ep_vrs}), we obtain the heliocentric semimajor axis of the asteroid at temporary capture as
\begin{eqnarray}
  \frac{a}{a_{\rm p}} \equiv \bar{a}_{\rm tc} = \Phi_{\rm p}\left[
    \frac{2}{A_\mp}-\left(\sqrt{2-\Phi_{\rm p}}\pm \sqrt{3}\nu\hat{r}_{\rm H}\right)^2
    \right]^{-1}.
  \label{eq:ep_aaa}
\end{eqnarray}
Note that Equation (\ref{eq:ep_aaa}) has four values 
corresponding to a combination of prograde or retrograde and $L_1$ or $L_2$.
If the sign in front of $\sqrt{3}$ is $"+"$, the temporary capture is prograde.
The $"-"$ sign represents retrograde capture.
The sign in $A_\mp$ represents $L_1$-type or $L_2$-type (Eq. (\ref{eq:ep_aa})).

\paragraph{Eccentricity.}
The heliocentric orbital angular momentum of the asteroid is 
\begin{eqnarray}
  h = rv\sin\alpha = \sqrt{GM_\odot a(1-e^2)}.
  \label{eq:ep_h1}
\end{eqnarray}
Substituting Equations (\ref{eq:ep_h1}) into condition [2] given by Eq.~(\ref{eq:ep_rr}) with $\alpha=\alpha_{\rm p}$, 
we obtain the heliocentric eccentricity at temporary capture,
\begin{eqnarray}
  e_{\rm tc}  = \sqrt{1-\sin\alpha_{\rm p}\left[1-\left(1-\frac{\Phi_{\rm p}A_\mp}{\bar{a}_{\rm tc}}\right)\right]}.
  \label{eq:ep_eee}
\end{eqnarray}

\paragraph{Angle of perihelion $\theta$.}
The perihelion angle at temporary capture
is easily obtained from Equation (\ref{eq:ep_rr0}), 
\begin{eqnarray}
  \theta_{\rm tc} &=& f_{\rm p} -{\rm acos}(g),
  \label{eq:ep_theta}
  \\
  g &=& \frac{\frac{\bar{a}_{\rm tc}}{\Phi_{\rm p}A_\mp}(1-e_{\rm tc}^2)-1}{e_{\rm tc}}
 = -e_{\rm tc}^{-1}\left(\cos^2\alpha_{\rm p}
  \mp \sin\alpha_{\rm p}\sqrt{e_{\rm tc}^2-\cos^2\alpha_{\rm p}}\right),
  \label{eq:ep_ggg}
\end{eqnarray}
where Equation (\ref{eq:ep_eee}) is substituted at the end.

\paragraph{Inclination.}

If the asteroid has non-zero heliocentric inclination $i$,
the relative velocity is modified.
Since the relative velocity is equal to $v_{\rm s}$, 
\begin{eqnarray}
  v_{\rm s}^2 &=& (v\cos i-v_{\rm p})^2 + v^2\sin^2 i
  \nonumber\\
  &=&v_{\rm p}^2\left[
    \left(\frac{2}{\Phi_{\rm p}A_\mp}-\frac{a_{\rm p}}{a}\right)+\left(\frac{2}{\Phi_{\rm p}}-1\right)
    -2\sqrt{\frac{2}{\Phi_{\rm p}A_\mp}-\frac{a_{\rm p}}{a}}\sqrt{\frac{2}{\Phi_{\rm p}}-1}\cos i
    \right]
  \nonumber\\
  &=&v_{\rm p}^2\left[\chi^2+\left(\frac{2}{\Phi_{\rm p}}-1\right)-2\chi\left(\frac{2}{\Phi_{\rm p}}-1\right)\cos i
    \right],
  \label{eq:ep_vr2}
\end{eqnarray}
which is reduced to
\begin{eqnarray}
  \chi^2+\left(\frac{2}{\Phi_{\rm p}}-1\right)-2\chi\sqrt{\frac{2}{\Phi_{\rm p}}-1}\cos i
  =(\nu v_{\rm H})^2.
\end{eqnarray}
For this equation to have a solution, 
the inclination must satisfy 
\begin{eqnarray}
\sin i <   \sqrt{\frac{3}{2-\Phi_{\rm p}}}\nu\hat{r}_{\rm H}.
  \label{eq:ep_imax}
\end{eqnarray}
The maximum value of $i$ for capture is obtained with $f_{\rm p}=180^\circ$ ($\Phi_{\rm p}=1+e_{\rm p}$).

\subsubsection{Dependence on $f_{\rm p}$ and $e_{\rm p}$}
\label{ss:ep_w}

\citet{hi16} found that capture is mostly retrograde for asteroids near the planetary orbit and is prograde for those from distant orbits. 
We found that this property does not change for a planet in an eccentric orbit.
The solutions to Equation (\ref{eq:ep_aaa}) and 
are plotted against $f_{\rm p}$ with $e_{\rm p}=0.2$ and a Jovian mass planet
for $\nu$ from $\nu=0$ (planetocentric circular orbit case) to
$\nu = \sqrt{2}$ (parabolic orbit cases) 
in Figures 1a and b. 
For $f_{\rm p} = 0$, the plot shows the following:
  \begin{eqnarray}
    \begin{array}{rclcl}
           &  \bar{a} & \la 0.6 & : &\;\;[\mbox{no capture},\;L_1]
  \nonumber\\
   0.6 \; (=\bar{a}_{\rm min})\; \la &\bar{a}& \la 0.8 & : &\;\;[\mbox{prograde},\;L_1]
    \nonumber\\
    0.8 \la &\bar{a}& \la 0.85 & : & \;\;[\mbox{prograde},\;L_1]
    \;\mbox{and}\;[\mbox{retrograde},\;L_2]
    \nonumber\\
    0.85 \la & \bar{a} & \la 1.2& : & \;\;[\mbox{retrograde},\;L_1, L_2] 
    \nonumber\\
    1.2 \la & \bar{a} & \la 1.45 & : & \;\;[\mbox{retrograde},\;L_1]
    \;\mbox{and}\;[\mbox{prograde},\;L_2]
    \nonumber\\
    1.45\la & \bar{a} & \la 2.9 \; (=\bar{a}_{\rm max})\; & : &\;\;[\mbox{prograde},\;L_2]
    \nonumber\\
    2.9 \la &  \bar{a} &  & : &\;\;[\mbox{no capture},\;L_1]
    \end{array}
  \end{eqnarray}
  The asteroids from these regions to the planet's Hill sphere have orbital eccentricities
  given by Eq.~(\ref{eq:ep_eee}).
  As seen in Figure 1a and b, 
  the boundaries of individual regions depend on $f_{\rm p}$.
  The planet can capture asteroids from further regions near perihelion ($f_{\rm p}$=0/360$^\circ$)
  than near aphelion.
  During a planet's orbital period, the instantaneous Hill radius $r_{\rm H}$ and
  $v_{\rm H}$ change.
  At its perihelion, $v_{\rm H}$ has the largest value, so that the planet captures
  asteroids from distant regions that have large relative velocity.
  Equation~(\ref{eq:ep_aaa}) suggests that the range of encounters, $\bar{a}_{\rm max}-\bar{a}_{\rm min}$, 
increases with $m_{\rm p}$ and $e_{\rm p}$, 
because $\hat{r}_{\rm H}\propto m_{\rm p}^{1/3}$ and $\Phi_{\rm s} \propto e_{\rm p}$ (for $e_{\rm p}^2 \ll 1$).
Numerically obtained values of $\bar{a}_{\rm max}$ and $\bar{a}_{\rm min}$ are plotted in Figure \ref{fig:ep_range}.

  Figure 1c and d show the solutions to Equation (\ref{eq:ep_ggg}) 
  with $\nu = \sqrt{2}$ for different values of $e_{\rm p}$.
  For $e_{\rm p}\sim 0$, $\theta_{\rm tc}$ covers all the range ($0^\circ$-$360^\circ$) 
  as $f_{\rm p}$ changes from $0^\circ$-$360^\circ$.
  The whole range is covered for small values of $\theta_{\rm tc}$ with slight modulation.
  However, for $e_{\rm p}$ larger than a threshold value ($e_{\rm p}^{\rm c}$),
  the coverage of $\theta_{\rm tc}$
  is only a part of $0^\circ$ to $360^\circ$.
  We will show that capture probability decreases with the increase in $e_{\rm p}$
  when $e_{\rm p} > e_{\rm p}^{\rm c}$.
  Since we found that $e_{\rm p}^{\rm c}$ is the largest for $\nu = \sqrt{2}$,
  we define the value for $\nu = \sqrt{2}$ with a given $m_{\rm p}$ as $e_{\rm p}^{\rm c}$ for $m_{\rm p}$.
 
\subsubsection{The dependence of $e_{\rm p}^{\rm c}$ on the planetary mass}
\label{ss:ep_ce}

  The values of $e_{\rm p}^{\rm c}$ are obtained numerically, 
  by finding if the point satisfying ${\rm d}\theta_{\rm tc}/{\rm d}f_{\rm p}=0$.
  Figure \ref{fig:ep_mp_cep} shows $e_{\rm p}^{\rm c}$
  for four types of temporary capture
  for
  $\nu=\sqrt{2}$
  as a function of $m_{\rm p}$.
  The dependence of $e_{\rm p}^{\rm c}$ on $m_{\rm p}$ is approximately given by
  $e_{\rm c}\simeq 5 \hat{r}_{\rm H} \propto m_{\rm p}^{1/3}$.
 
In the figure, the current values of the eccentricities of the eight planets of
  the solar system
  are also plotted.
  The bars attached to the points show the maximum variation ranges over past 10 Myr,
  calculated by following the method developed by \citet{i95} which
  is based on the secular perturbation theory of \citet{l88}.
As we will show later, the analytically derived values of $e_{\rm p}$,
beyond which the temporary capture probability drops, agree with the results obtained by
numerical orbital integration.
  Jupiter, Saturn, and Neptune always have $e_{\rm p}<e_{\rm p}^{\rm c}$.
  This means that their rates of temporary capture have remained relatively high.
  The maximum $e_{\rm p}$ values for Venus, Earth, and Uranus are slightly higher than
  $e_{\rm p}^{\rm c}$ but the current values and most of the error-bar ranges
  of $e_{\rm p}$ are below $e_{\rm p}^{\rm c}$.

  Mars, which has relatively high $e_{\rm p}$, apparently has less chance to
  capture asteroids with its current orbit.
  However, the bar of $e_{\rm p}$ for Mars shows that the Martian $e_{\rm p}$ 
  can have the values of $e_{\rm p}$ much smaller than $e_{\rm p}^{\rm c}$ during
  orbital variations.
  Mercury never has $e_{\rm p}<e_{\rm p}^{\rm c}$.

\section{THE EFFICIENCY OF TEMPORARY CAPTURE BY AN ECCENTRIC PLANET}
\label{ss:ep_kkk}

Now we estimate the dependences of the probability of temporary capture 
on $e_{\rm p}$ and $m_{\rm p}$ of the host planet.
We define the probability as $K_{\rm tc}/K_{\rm enc}$,
  where $K_{\rm enc}$ and $K_{\rm tc}$ are
  the phase space volume that satisfies the conditions
  for encounters with the planet's Hill sphere, and
  that for temporary capture, respectively.
Encounters wth the Hill sphere are defined as those
with minimum distance to the planet less than 
their instantaneous Hill radius $r_{\rm H}$.
For simplicity, we here set $r_{\rm p}=a_{\rm p}(1+e_{\rm p}^2/2)$.
For example, we consider a close encounter orbit with $\bar{a}<1$. 
The maximum eccentricity $e_1$ is required for an orbit with its aphelion at the $L_2$ point;
\begin{eqnarray}
  \bar{a}(1+e_1)=\left(1+\frac{1}{2}e_{\rm p}^2\right)
  \left(
  1+\hat{r}_{\rm H}\right)&\;\;\rightarrow\;\;&1+e_1=\left(1+\frac{1}{2}e_{\rm p}^2\right)
  \left(\frac{1+\hat{r}_{\rm H}}{\bar{a}}\right).
\end{eqnarray}
In a similar way, the minimum eccentricity $e_2$ satisfies
\begin{eqnarray}
  \bar{a}(1+e_2)=\left(1+\frac{1}{2}e_{\rm p}^2\right)\left(
  1-\hat{r}_{\rm H}\right)&\;\;\rightarrow\;\;&1+e_2=
  \left(1+\frac{1}{2}e_{\rm p}^2\right)\left(\frac{1-\hat{r}_{\rm H}}{\bar{a}}\right).
\end{eqnarray}
Then, the range of eccentricity for close encounters is given by
\begin{eqnarray}
  \Delta e = e_1-e_2 = \left(1+\frac{1}{2}e_{\rm p}^2\right)\frac{2\hat{r}_{\rm H}}{\bar{a}}.
\end{eqnarray}
The range of eccentricity for close encounters with $\bar{a}>1$ is the same. 
The range of the angle of perihelion for close encounters is $\Delta\theta/2\pi$,
where we can set $\Delta\theta=2\hat{r}_{\rm H}$.
Then we obtain $K_{\rm enc}$ as the phase space volume by integrating 
  $\Delta e \cdot \Delta\theta/2\pi$
  over $\bar{a}$ with the time weight ($\propto \bar{a}^{-3/2}$), 
\begin{eqnarray}
  K_{\rm enc} = \frac{\hat{r}_{\rm H}}{\pi}
  \int_{\bar{a}_{\rm min}}^{\bar{a}_{\rm max}}
  \Delta e(\bar{a}) \, \bar{a}^{-\frac{3}{2}}{\rm d}\bar{a}
  = \frac{2 \hat{r}_{\rm H}^2}{3\pi}\left(1+\frac{1}{2}e_{\rm p}^2\right)
  \left(\bar{a}_{\rm min}^{-\frac{3}{2}}-\bar{a}_{\rm max}^{-\frac{3}{2}}\right),
  \label{eq:ep_kkkenc}
\end{eqnarray}
where we assumed a uniform $a$-distribution of asteroids. 
We use $\bar{a}_{{\rm tc, min}, L_1}$ and $\bar{a}_{{\rm tc, max}, L_2}$
for $\bar{a}_{\rm min}$ and $\bar{a}_{\rm max}$,
which are obtained from Equation (\ref{eq:ep_aaa}).
We set the upper limit of $\bar{a}_{\rm max}=3$ 
to avoid the divergence in the calculation of $K_{\rm enc}$.
This is used only in cases of Jovian mass planets.
Assuming $e_{\rm p}\ll1$ and $\hat{r}_{\rm H}\ll1$, one can find that $K_{\rm enc}\propto\hat{r}_{\rm H}^3$.
  
The phase space volume for temporary capture is much more restricted than for the encounters.
In a similar way as we defined $K_{\rm enc}$,
the phase volume of temporary capture is given by  
   \begin{eqnarray}
     K_{\rm tc} = \frac{1}{T_{\rm p}}
    \int_0^{T_{\rm p}} \int_{\bar{a}_{\rm min}}^{\bar{a}_{\rm max}}
    \Delta e_{\rm tc} \, \frac{\Delta \theta_{\rm tc}}{2\pi} \, \bar{a}_{\rm tc}^{-\frac{3}{2}}{\rm d}\bar{a}_{\rm tc}{\rm d}t.
    \label{eq:Ktc0}
    \end{eqnarray}
  Because $a_{\rm tc}$, $e_{\rm tc}$, and $\theta_{\rm tc}$ are correlated,
  it is useful to rewrite $\Delta e_{\rm tc}$, $\Delta \theta_{\rm tc}$, and 
  ${\rm d} a_{\rm tc}$ as $\Delta e_{\rm tc}=({\rm d} e_{\rm tc}/{\rm d} \nu_{\rm tc})\Delta \nu_{\rm tc}$, 
  $\Delta \theta_{\rm tc}=({\rm d} \theta_{\rm tc}/{\rm d} \nu_{\rm tc})\hat{r}_{\rm H}\Delta \nu_{\rm tc}$,
  and ${\rm d} \bar{a}_{\rm tc}=({\rm d} \bar{a}_{\rm tc}/{\rm d} \nu_{\rm tc})d \nu_{\rm tc}$.
  Using these relations, we change the integral of $K_{\rm tc}$
  by ${\rm d}a_{\rm tc}$ to that by $\Delta \nu_{\rm tc}$. 
  For $e_{\rm p}=0$, we set $\Delta \theta_{\rm tc}=\Delta \gamma \cdot \hat{r}_{\rm H}$,
  where $\Delta \gamma \ll 1$.
  Because the integrands depend on $f_{\rm p}$, we also added time averaging over 
  an orbital period of the planet ($T_{\rm p}=1$).

  Thereby, the temporary capture rate is given from Equations (\ref{eq:ep_aaa}) as
  \begin{eqnarray}
     K_{\rm tc} & = & \frac{(\Delta \nu)^2\hat{r}_{\rm H}}{2\pi T_{\rm p}}
    \int_0^{T_{\rm p}} \int_{\nu_{\rm min}}^{\nu_{\rm max}}
    \frac{{\rm d} e_{\rm tc}}{{\rm d} \nu_{\rm tc}}
    \frac{{\rm d} \theta_{\rm tc}}{{\rm d} \nu_{\rm tc}}
    \frac{{\rm d}\bar{a}_{\rm tc}}{{\rm d} \nu_{\rm tc}}
    \, \bar{a}_{\rm tc}^{-\frac{3}{2}}{\rm d}\nu_{\rm tc}{\rm d}t,
    \label{eq:Ktc}\\
    \frac{{\rm d }\bar{a}_{\rm tc}}{{\rm d} \nu}
    &=&
    \pm 2\frac{\bar{a}_{\rm tc}^2}{\Phi_{\rm p}}\sqrt{3}\hat{r}_{\rm H}
    \left(\sqrt{2-\Phi_{\rm p}}\pm \sqrt{3}\hat{r}_{\rm H}\nu\right)
    \label{eq:ep_dadn}
    \\
    \frac{{\rm d}  e_{\rm tc}}{{\rm d} \nu}
    &=& \frac{{\rm d}  e_{\rm tc}}{{\rm d} \bar{a}_{\rm tc}}
    \frac{{\rm d }\bar{a}_{\rm tc}}{{\rm d} \nu}
    \label{eq:ep_dedn}
    \\
    \frac{{\rm d}  \theta_{\rm tc}}{{\rm d} \nu}
    &=& \frac{\frac{{\rm d}g}{{\rm d}\nu}}{\sqrt{1-g^2}}
    \label{eq:ep_dthdn} \\
    \frac{{\rm d}  e_{\rm tc}}{{\rm d}  \bar{a}_{\rm tc}}
    & = &
    \Phi_{\rm p}A_\mp\sin^2\alpha_{\rm p}
    \left(1-\frac{\Phi_{\rm p}A_\mp}{\bar{a}_{\rm tc}}\right)
    \bar{a}_{\rm tc}^{-2}e_{\rm tc}^{-1}
    \label{eq:ep_deda} \\
    \frac{{\rm d}  g}{{\rm d}  \nu} & = &
    \frac{{\rm d}  g}{{\rm d} e_{\rm tc}}\frac{{\rm d}e_{\rm tc}}{{\rm d}\nu}; \;
    \frac{{\rm d}  g}{{\rm d}  e_{\rm tc}}
    =  -ge_{\rm tc}^{-1}
    \pm\frac{\sin\alpha_{\rm p}}{\sqrt{e_{\rm tc}^2-\cos^2\alpha_{\rm p}}}.
    \label{eq:ep_dgde}
  \end{eqnarray}
For $e_{\rm p}=0$, 
\begin{eqnarray}
  K_{\rm tc} & = & \frac{\Delta \nu\Delta\gamma\hat{r}_{\rm H}}{2\pi}
    \int_{\nu_{\rm min}}^{\nu_{\rm max}}
    \frac{{\rm d} e_{\rm tc}}{{\rm d} \nu_{\rm tc}}
    \frac{{\rm d}\bar{a}_{\rm tc}}{{\rm d} \nu_{\rm tc}}
    \, \bar{a}_{\rm tc}^{-\frac{3}{2}}{\rm d}\nu_{\rm tc}
    \label{eq:Ktc1d}
\end{eqnarray}
Assuming $e_{\rm p}\ll1$ and $\hat{r}_{\rm H}\ll1$,
  one can find
  $({\rm d} a_{\rm tc}/{\rm d} \nu_{\rm tc})\propto\hat{r}_{\rm H}$,
  and 
  $({\rm d} e_{\rm tc}/{\rm d} \bar{a}_{\rm tc})$,
  $({\rm d} g/{\rm d} e_{\rm tc})$, and $g$ are independent of $\hat{r}_{\rm H}$.
  This leads to $K_{\rm tc}\propto\hat{r}_{\rm H}^3$,
  which is the same as $K_{\rm enc}$, implying that $K_{\rm tc}/K_{\rm enc}$
  is independent of $m_{\rm p}$ for 
  $e_{\rm p} \ll 1$.

    The integration range, $\nu_{\rm min}<\nu_{\rm tc}<\nu_{\rm max}$, can be
    simply estimated in the framework of the two-body problem
    (planet-particle) as follows.
  The physical radius of the planet may give the value of $\nu_{\rm min}$.
  A planetocentric temporarily
  captured orbit has its apocenter distance at $a_{\rm s}(1+e_{\rm s}) \simeq r_{\rm H}$.
  The pericenter distance, $a_{\rm s}(1-e_{\rm s})$,
  must be larger than the physical radius of the planet, $R_{\rm p}$, to avoid a collision.
  From these two equations, 
  \begin{eqnarray}
    e_{\rm s} < \frac{1-(R_{\rm p}/r_{\rm H})}{1+(R_{\rm p}/r_{\rm H})}
  \end{eqnarray}
  Since $\nu=\sqrt{\kappa}=\sqrt{1-e_{\rm s}}$ for $f_{\rm s}=0$,
  \begin{eqnarray}
   \nu_{\rm min} = \sqrt{\frac{R_{\rm p}/r_{\rm H}}{1+(R_{\rm p}/r_{\rm H})}}\simeq\sqrt{R_{\rm p}/r_{\rm H}}.
    \label{eq:ep_numin}
  \end{eqnarray}
  The simplest assumption for the maximum value
    in the framework of the two-body problem
  is $\nu_{\rm max}=\sqrt{2}$,
  which is the upper limit
  for an elliptic orbit around the planet.
    However, the effect of the third body (Sun) changes them.
    We found that it is more appropriate
    to assume $\nu_{\rm min} = 0.25$ and $\nu_{\rm max} = 2$
    for a prograde trap and $\nu_{\rm min} = 0.5$ and $\nu_{\rm max}= \sqrt{2}$
    for a retrograde trap from the orbital calculations described below.
    We adopt these integration ranges.
  
Figure \ref{fig:ep_ep_k_ana} shows the ratio $K_{\rm tc}/K_{\rm enc}$ as a function
of $e_{\rm p}$ for planets with Martian, Jovian, Earth, and Neptunian mass.
Each plot has four curves for the temporary capture types, 
and the sum of the four types (the black line).
The total ratio (black) is almost
constant or rather gradually increases with $e_{\rm p}$.
until $e_{\rm p}$ exceeds $e_{\rm p}^{\rm c}$.
The asymptotic values of $K_{\rm tc}/K_{\rm enc}$ at $e_{\rm p}\rightarrow 0$
  are independent of planetary mass ($m_{\rm p}$) and semimajor axis ($a_{\rm p}$),
  as we predicted.
As shown in Figure \ref{fig:ep_ep_k_ana},
    $e_{\rm p}^{\rm c}\simeq 5 \hat{r}_{\rm H}$ where 
    $\hat{r}_{\rm H}=4.8\times 10^{-3}$ for Mars,
    $\hat{r}_{\rm H}=0.068$ for Jupiter,
    $\hat{r}_{\rm H}=0.01$ for Earth, and $\hat{r}_{\rm H}=0.026$ for Neptune.
For $e_{\rm p} > e_{\rm p}^{\rm c}$s, $K_{\rm tc}$
decays with $e_{\rm p}$ approximately as $\propto e_{\rm p}^{-1}$.
As will be shown in the next section,
the functional form of the predicted $K_{\rm tc}/K_{\rm enc}$ 
agrees very well with the results of numerical orbital integrations,
while the allowance for temporary capture $\Delta \nu$ and $\Delta\gamma$
 cannot be estimated by analytical arguments here.
Because $\nu$ expresses the satellite orbital energy at the Hill radius, it is expected
that the allowance $\Delta \nu$ is independent of $m_{\rm p}$ and $a_{\rm p}$ as well.
Also the independence of $\Delta\gamma$ is expected
  since the angle $\Delta\theta_{\rm tc}$ would be a function
  only of $\hat{r}_{\rm H}$.
From comparison with the numerical simulations,
we empirically set
$\Delta \nu \sim 0.025$ and $\Delta \gamma \sim 0.05$.

\section{Comparison with Numerical Results}
\label{ss:ep_nu}
We perform numerical calculations for the temporary capture of bodies
by planets with Mars, Jupiter, Earth, and Neptune masses
to evaluate the relevance of our analytical formulae.

\subsection{Methods and Initial Conditions}

We compute the orbital evolution of massless bodies, which correspond to asteroids, perturbed by
  a planet in a circular or eccentric orbit,
  using a 4th-order Hermite integration scheme.
  The parameters are summarized in Table \ref{tb:ep_p}.
  The number of the massless bodies in each run is $5\times 10^6$.
  Asteroids are initially uniformly distributed on the
  $a, e$-plane between $\bar{a}_{{\rm tc, min}, L_1}<\bar{a}<\bar{a}_{{\rm tc, max}, L_2}$,
  $e_{\rm min}<e<e_{\rm max}$, 
  which are derived analytically and numerically in Section \ref{ss:ep_w} and
  summarized in Figure \ref{fig:ep_range}.
  The parameter $\theta$ is also uniformly distributed between 0 and $2\pi$.
  We set the upper limit of $\bar{a}_{{\rm tc, max}, L_2}=3$.
  In most runs we assume $i=0$ for the asteroids.
  In several additional runs, 
  we give $i$ with a uniform distribution for $0<i<i_{\rm tc, max}$
  where $i_{\rm tc, max}$ is given by Equation (\ref{eq:ep_imax})
  for $f_{\rm p}=180^\circ$ and $\nu=\sqrt{2}$.
We regard asteroids as temporarily captured bodies
  if they stay within $r_{\rm H}$ from the planet
  longer than one orbital period of the planet
  $T_{\rm p}$.
 
  Using the planetocentric location and the relative velocity vector to the planet,
   at the moment when an asteroid enters the $r_{\rm H}$ region 
  around the planet for the first time, 
  we define the type of temporary capture:
  [prograde-$L_1$], [retrograde-$L_1$], [retrograde-$L_2$], and [prograde-$L_2$].
    
  In this paper, we focus on the equilibrium state 
  where the ratio of temporary capture and encounter rates
  becomes constant with time.
  To obtain this state, we first perform several long-time calculations
  with $10^5$ particles for $10^5T_{\rm p}$ and 
 choose the time range where the ratio
  is constant with time.
  Note that \citet{hi16} presented the cumulative number of captured bodies  
  over $10^6$ years, which is not directly compared with 
  the results presented here.

\subsection{Results}

Figure \ref{fig:ep_fp_theta}
shows $\theta$ of the temporarily captured bodies 
against $f_{\rm p}$ for a Martian mass planet with various $e_{\rm p}$.
The analytical prediction (eq. (\ref{eq:ep_theta})) 
for $0\le\nu\le\sqrt{2}$ is also plotted.
The analytical prediction,
which determines the critical eccentricity for temporary capture ($e_{\rm p}^{\rm c}$), 
agrees well with the numerical results.

Figure \ref{fig:ep_ep_k_num} shows
the ratio ($n_{\rm tc}/n_{\rm enc}$) of the temporary capture and 
encounter rates as a function of $e_{\rm p}$ for planets with
Martian, Jovian, Earth, and Neptunian mass, respectively.
The ratio drops beyond the predicted values of $e_{\rm p}^{\rm c} \simeq 5 \hat{r}_{\rm H}$,
which are 0.02, 0.27, 0.04, and 0.1 for  Martian, Jovian, Earth, and Neptunian mass.
  This drop of $n_{\rm tc}/n_{\rm enc}$ is well reproduced by the analytical
  prediction in Figure \ref{fig:ep_ep_k_ana}.

The value of $n_{\rm tc}/n_{\rm enc}$ for $e_{\rm p}<e_{\rm p}^{\rm c}$
  is $\sim 10^{-3}$, which is almost independent of the planetary mass, as predicted.
  We performed additional numerical calculations using particles
  with $i<i_{\rm max}$ for $\nu=\sqrt{2}$ given by Equation (\ref{eq:ep_imax}).
  The results
  show that the values of $n_{\rm tc}/n_{\rm enc}$ for the 3D calculations
  are similar to those for the 2D calculations (within a factor of 2).

\section{SUMMARY AND DISCUSSION}

In order to explore the origins of irregular or minor satellites around the planets
in the solar system, we have investigated the probability of temporary capture
through semi-analytical arguments and numerical integration.
We extended the analysis of temporary capture around a planet in a circular orbit
developed by \citet{hi16} to that around a planet in an eccentric orbit,
allowing us to discuss the origins of the Martian satellites.
We derived the capture probability as a function of planetary mass ($m_{\rm p}$) 
and eccentricity ($e_{\rm p}$).
Analytical formulae reproduce the numerical integrations very well.

We found that the temporary capture
occurs at $\sim 0.1\%$ of encounters that enter Hill sphere of a planet,
independent of $m_{\rm p}$, $a_{\rm p}$ (semimajor axis) and $e_{\rm p}$
up to a critical value $e_{\rm p}^{\rm c} \simeq 5 (m_{\rm p}/3M_\odot)^{1/3}$.
For $e_{\rm p} > e_{\rm p}^{\rm c}$, the probability decays with increasing $e_{\rm p}$ as
$\propto e_{\rm p}^{-(1-2)}$.

The current eccentricity of Mars is several times larger than $e_{\rm p}^{\rm c}$,
so that the capture origin of Phobos and Deimos looks unfavored.
However, as shown in Figure \ref{fig:ep_mp_cep}, the Martian eccentricity changes
with time and can be lower than $e_{\rm p}^{\rm c}$ for some fraction of time,
and temporary capture may have been available in the past.
  Note again that temporary capture is a necessary condition for permanent
capture and their respective probabilities are not necessarily proportional to each other.
As will be discussed in a separate paper, tight capture could be found
in the cases where $e_{\rm p} > e_{\rm p}^{\rm c}$.
In a subsequent paper, we will
  discuss the probability of permanent capture and
the possibility of the capture origin of Phobos and Deimos.

\acknowledgements
We thank an anonymous referee for his/her useful comments that helped to improve the paper.
This work was supported by JSPS KAKENHI grant Number 23740335 and 15H02065.
Data analyses were in part carried out on the PC cluster at 
the Center for Computational Astrophysics, 
National Astronomical Observatory of Japan.

\begin{table}[hbtp]
  \begin{center}
    \scalebox{0.8}{
      \begin{tabular}{c|ccc}
	\hline\hline
	Planet&$a_{\rm p}$ (au)&$m_{\rm p}$ ($M_\odot$)& $e_{\rm p}$ Range\\
	\hline
        Earth  & 1  & 3.00e$-$06 & 0.004-0.36\\
        \hline
        Mars  & 1.52  & 3.72e$-$07 & 0.002-0.18\\
        \hline
        Jupiter  & 5.2  & 9.55e$-$04  & 0.01-0.9\\
        \hline
        Neptune  & 30.1  & 5.15e$-$05  & 0.005-0.5\\
        \hline
      \end{tabular}
    }
    \caption{Parameters of planets Used in Numerical Calculations.}
    \label{tb:ep_p}
  \end{center}
\end{table}

\begin{figure}[hbtp]
  \begin{center}
    \resizebox{13cm}{!}{\includegraphics{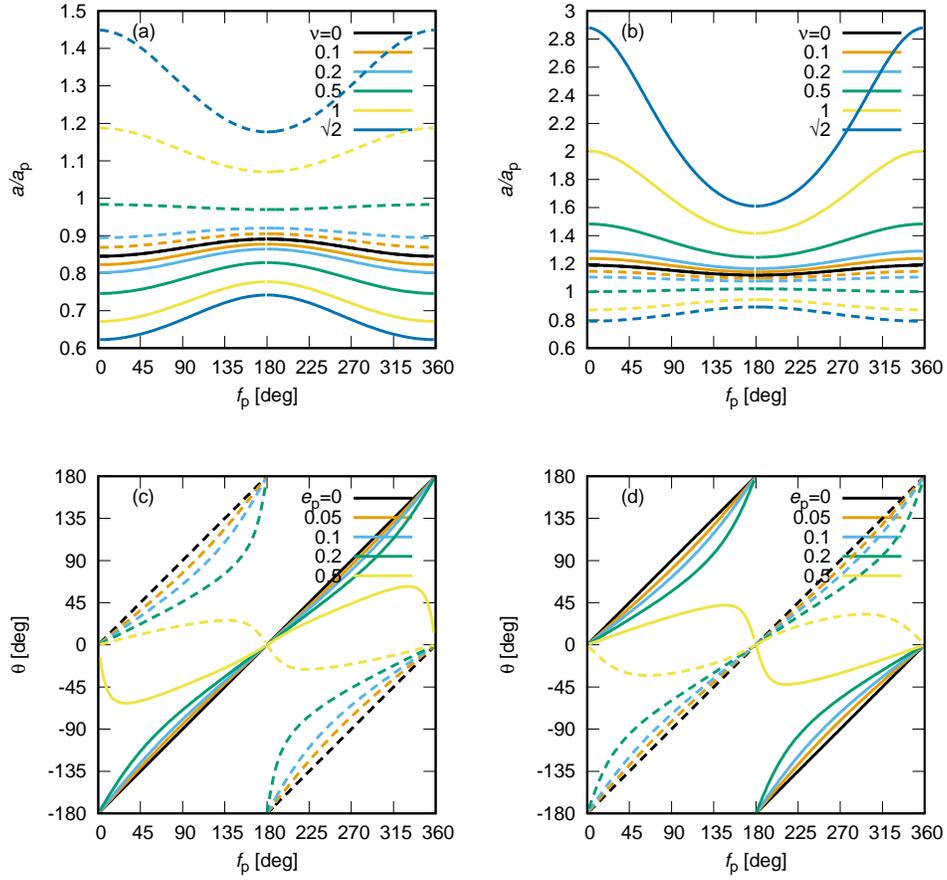}}
    \caption{
      The solutions to Equations (\ref{eq:ep_aaa}) are
      plotted against $f_{\rm p}$ with $e_{\rm p}=0.2$ and a Jovian mass planet
      for $\nu=0$(black), 0.1 (orange), 0.2 (light blue), 0.5 (green), 1 (yellow), and $\sqrt{2}$ (blue): 
      (a) $L_1$-type and (b) $L_2$-type captures.
      The solutions to Equations (\ref{eq:ep_theta}) with $\nu=\sqrt{2}$ and a Jovian mass planet
      are plotted for $e_{\rm p}=0$, 0.01, 0.02, 0.05, and 0.09: (c) $L_1$-type and (d) $L_2$-type captures.
      The solid and dashed curves are for prograde and retrograde captures, respectively.
    }
    \label{fig:ep_fp_oeJ}
  \end{center}
\end{figure}

\begin{figure}[hbtp]
  \begin{center}
   \resizebox{13cm}{!}{\includegraphics{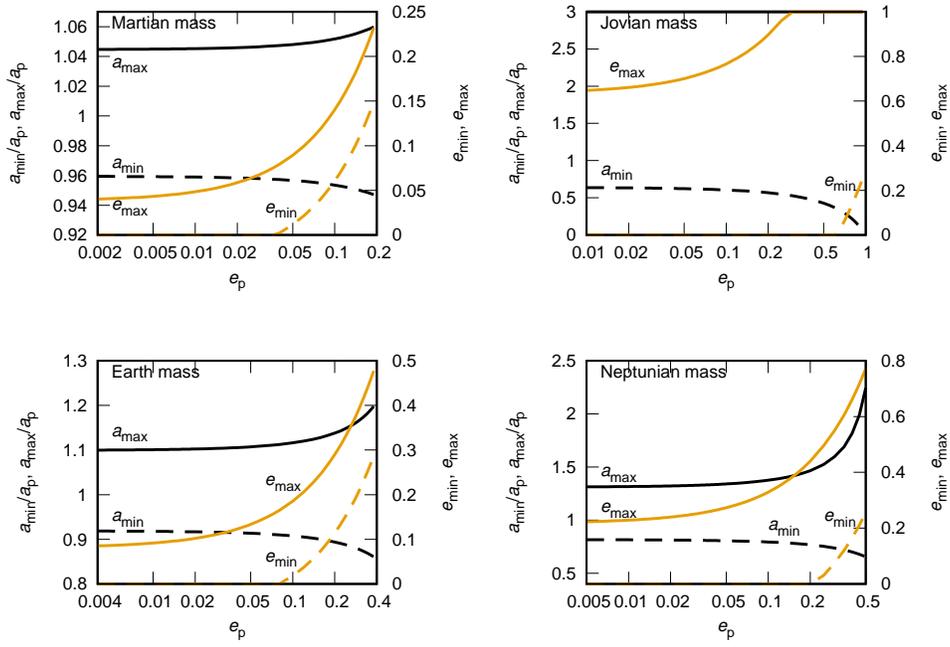}}
    \caption{
      Ranges of initial orbital elements are summarized for each planetary mass
      against $e_{\rm p}$
      (top-left: Martian mass, top-right: Jovian mass, bottom-left: Earth mass,
      bottom-right: Neptunian mass.)
      Black curves show $\bar{a}_{\rm max}$(solid) and $\bar{a}_{\rm min}$(dashed)
      on the left $y-$axis and 
      orange curves show $e_{\rm max}$(solid) and $e_{\rm min}$(dashed)
      on the right $y-$axis.
    }
    \label{fig:ep_range}
  \end{center}
\end{figure}

\begin{figure}[hbtp]
  \begin{center}
   \resizebox{13cm}{!}{\includegraphics{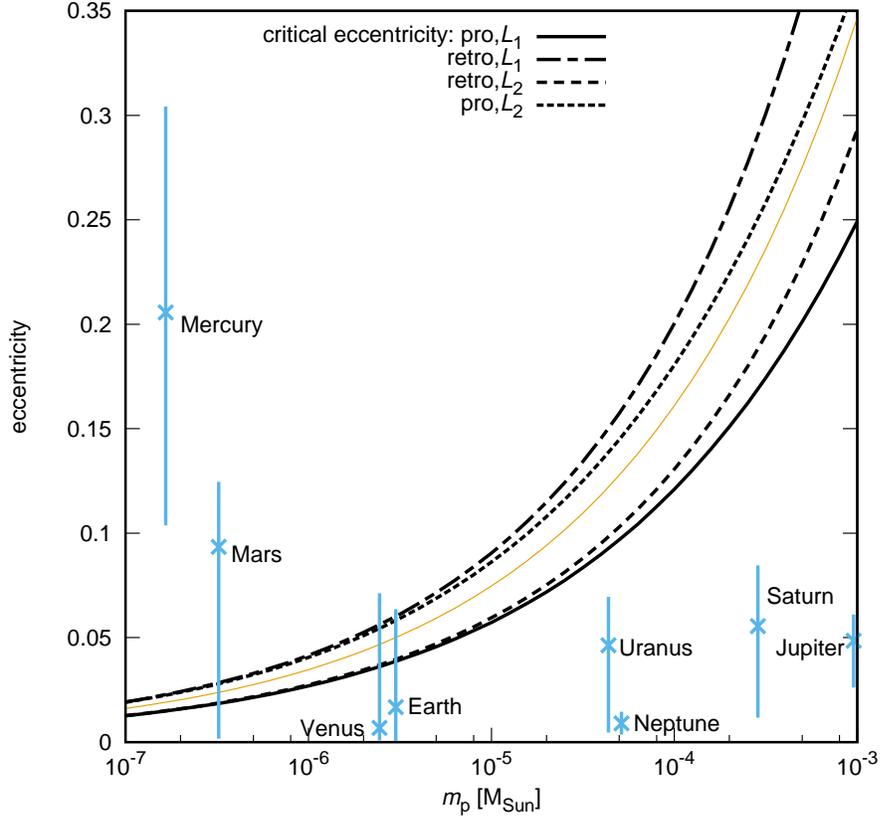}}
    \caption{
      Critical eccentricity $e_{\rm p}^{\rm c}$ for $\nu=\sqrt{2}$ is plotted against $m_{\rm p}$.
      The curve types indicate the capture type 
      (solid: [prograde, $L_1$], long-short dashed: [retrograde, $L_1$],
      dashed: [retrograde, $L_2$], short dashed: [prograde, $L_2$].)
      The orange curve shows $e_{\rm p}=5\hat{r}_{\rm H}$.
      Current eccentricities of eight planets of the solar system are also plotted against their mass.
      The error bars show the variations over 10 Myr calculated following \citet{l88}.
    }
    \label{fig:ep_mp_cep}
  \end{center}
\end{figure}

\begin{figure}[hbtp]
  \begin{center}
    \resizebox{13cm}{!}{\includegraphics{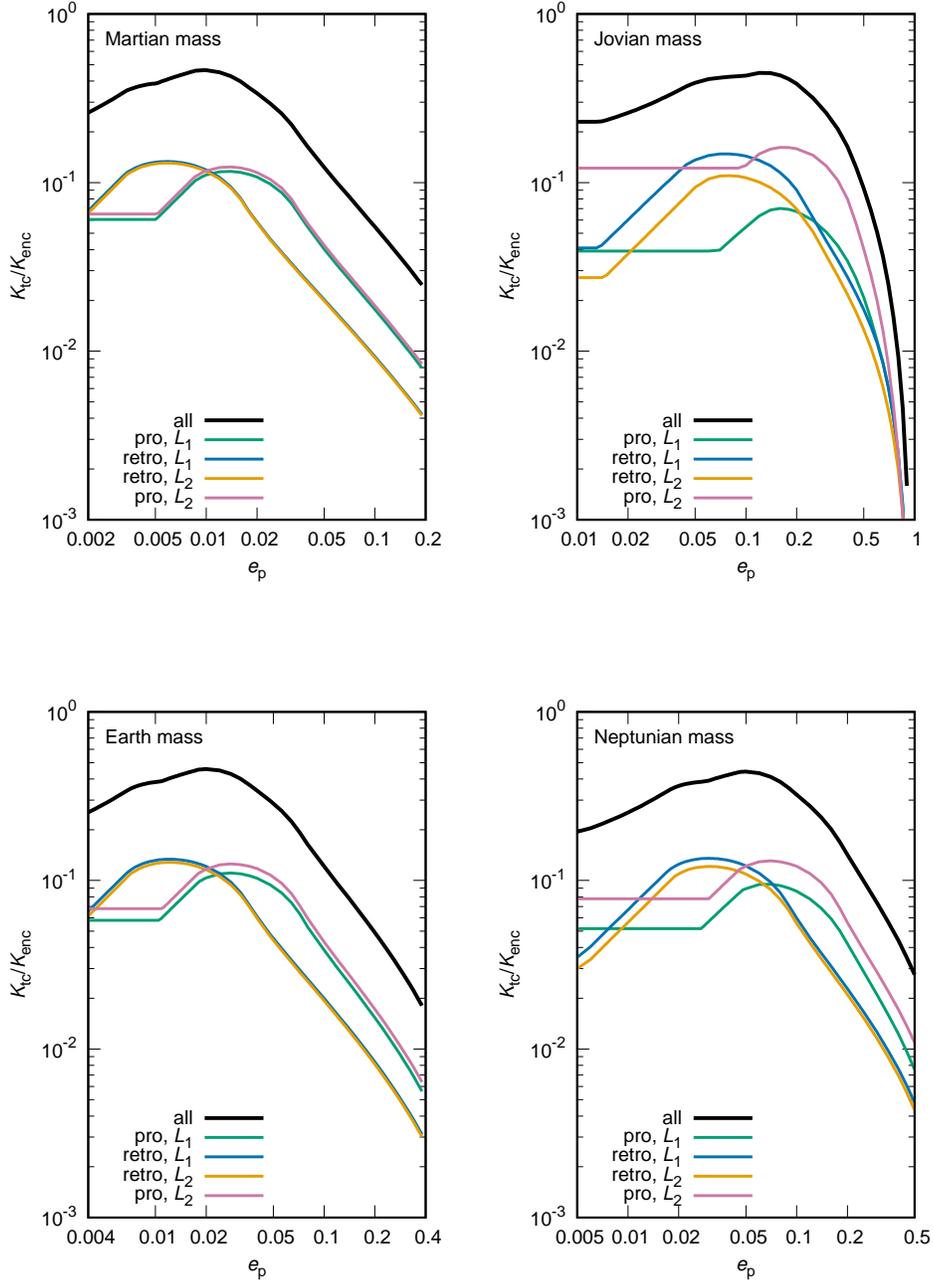}}
    \caption{
      Efficiency of temporary capture $K_{\rm tc}/K_{\rm enc}$
      plotted against $e_{\rm p}$
      for planets with Martian (top-left),
      Jovian (top-right), Earth (bottom-left), and Neptunian (bottom-right)
      mass using $\nu_{\rm min}=0.25$ and $\nu_{\rm max}=2$ for prograde
      and $\nu_{\rm min}=0.5$ and $\nu_{\rm max}=\sqrt{2}$ for retrograde.
      We set $\Delta \nu \sim 0.025$ and $\Delta \gamma \sim 0.05$.
      The $K_{\rm tc}$
      for each temporary capture type is plotted in color;
      [prograde, $L_1$] (green), [retrograde, $L_1$] (blue),
      [retrograde, $L_2$] (orange), and [prograde, $L_2$] (pink).
      The black curve shows the sum of the four types.
      We adopt Equation (\ref{eq:Ktc1d}) for $K_{\rm tc}$ 
      if $K_{\rm tc}$ with Equation (\ref{eq:Ktc})
      for $e_{\rm p}<e_{\rm p}^{\rm c}$ is less than
      that with Equation (\ref{eq:Ktc1d}).
    }
    \label{fig:ep_ep_k_ana}
  \end{center}
\end{figure}

\begin{figure}[hbtp]
  \begin{center}
    \resizebox{13cm}{!}{\includegraphics{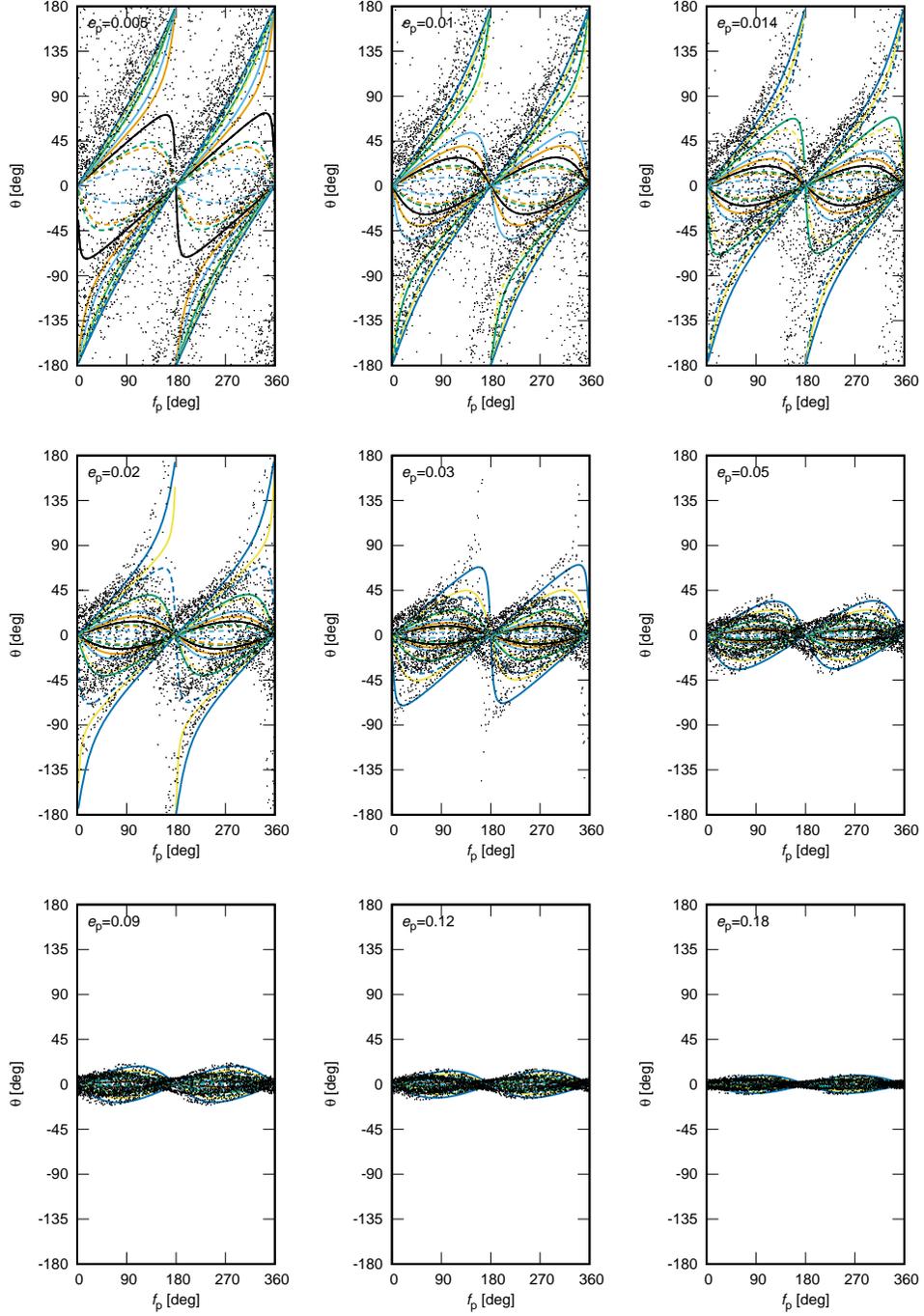}}
    \caption{
      Argument of perihelion of temporarily captured bodies
      by a Martian mass planet with various $e_{\rm p}$
      at the moment of entering the Hill sphere for the first time
      are plotted against $f_{\rm p}$.
      The solution to Equation (\ref{eq:ep_theta})
      for each temporary capture type
      for $\nu=0$ (black), 0.1 (orange), 0.2 (light blue), 
      0.5 (green), 1 (yellow), and
      $\sqrt{2}$ (blue)
      are also plotted.
      Solid and dashed curves are for prograde and retrograde temporary capture, respectively.
      All types of temporary capture are plotted in the same panel.
    }
    \label{fig:ep_fp_theta}
  \end{center}
\end{figure}

\begin{figure}[hbtp]
  \begin{center}
    \resizebox{13cm}{!}{\includegraphics{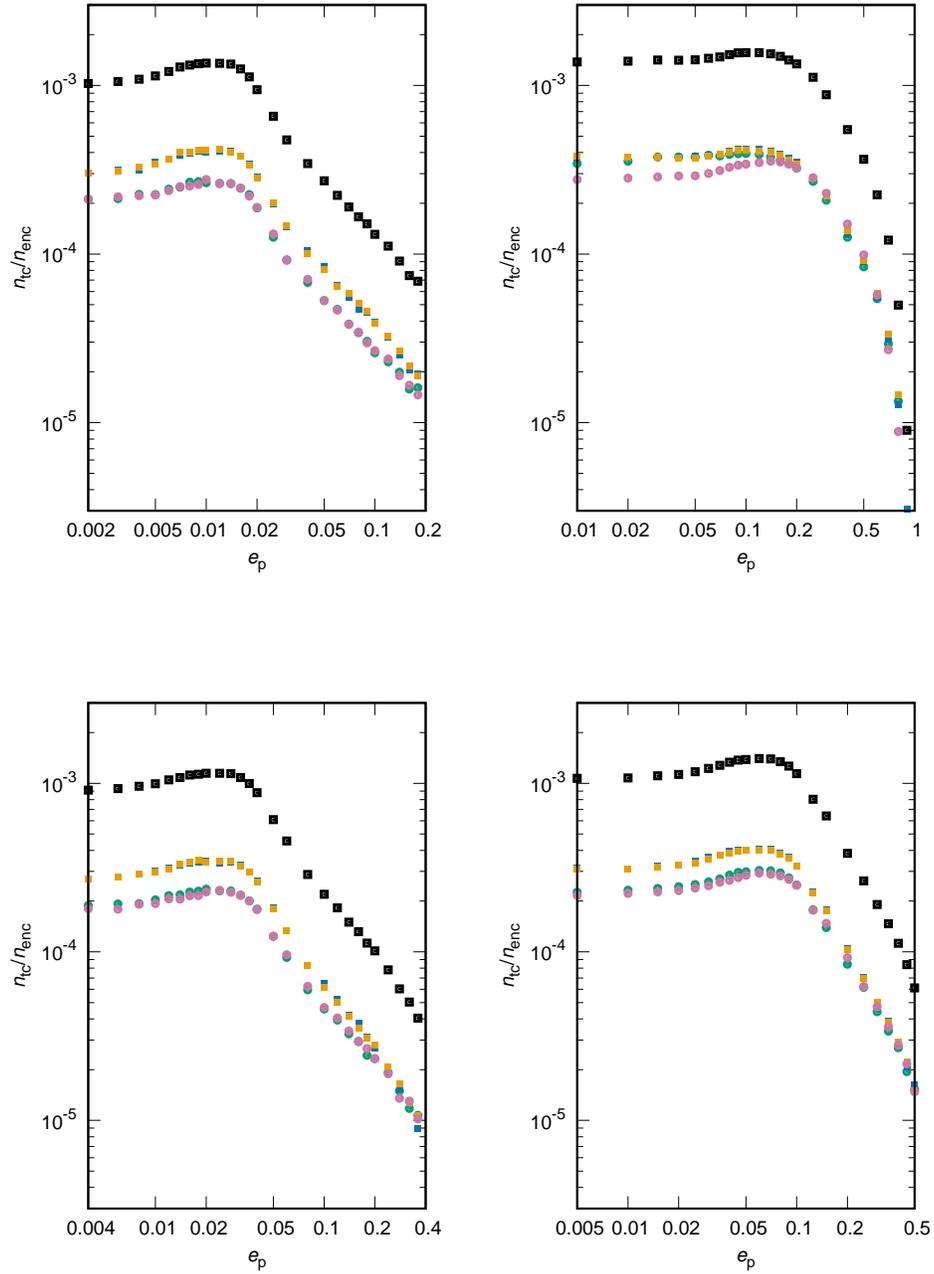}}
    \caption{
      The ratio of the number of temporary captures $n_{\rm tc}$
      to that of encounters $n_{\rm enc}$ is plotted
      against $e_{\rm p}$
      for planets with Martian (top-left),
      Jovian (top-right), Earth (bottom-left), and Neptunian (bottom-right)
      mass.
      The colors indicate the types of temporary capture;
      [prograde, $L_1$] (green), [retrograde, $L_1$] (blue),
      [retrograde, $L_2$] (orange), [prograde, $L_2$] (pink),
      and the sum of all types(black).
    }
    \label{fig:ep_ep_k_num}
  \end{center}
\end{figure}



\end{document}